\input{aipcheck}

\documentclass[
    ,final            
  ]
  {aipproc}

\layoutstyle{8x11single}

\begin{document}

\title{The Blazar Spectral Sequence and GLAST}


\author{L. Maraschi, G. Ghisellini and F. Tavecchio}
{address={INAF-Osservatorio Astronomico di Brera, Milano, Italy}
}

\begin{abstract}
The present status and understanding of the "spectral sequence"
of blazars is discussed in the perspective of the upcoming GLAST 
launch. The vast improvement in sensitivity will allow to 
i) determine more objectively the "average" gamma-ray properties
of classes  objects ii) probe more deeply the ratio between 
accretion power and jet power in different systems.
\end{abstract}

\maketitle


\section{Introduction}

The spectral sequence of blazars (Fossati et al.,1998 al.,1998) was
constructed merging three complete blazar samples (two radio selected,
one X-ray selected: 2 Jy FSRQ, Wall \& Peacock 1985, 1Jy BL Lac, Kuhr
et al. 1981, and Slew Survey BL Lac, Elvis et al. 1992), grouping all
the objects in radio luminosity bins and averaging monochromatic
luminosities of objects within each radio-luminosity bin. The
procedure is thus prone to various biases (Maraschi \& Tavecchio
2001), in particular the gamma-ray data were largely incomplete.

The resulting "sequence" shows that the blazar SEDs are double humped
and that the two peaks shift to higher energies with decreasing
luminosity.  Systematic modelling of the SEDs of individual objects
(Ghisellini et al. 1998) yields basically uniform beaming factors and
jet parameters varying along the sequence in the sense of an
increasing energy density and decreasing electron critical energy at
higher luminosities.  Thus the "sequence" offers a suggestive
indication that the basic spectral properties of blazar jets could be
related to the different powers involved and possibly represent an
evolutionary sequence in cosmic history (Boettcher and Dermer 2002;
Cavaliere and D'Elia 2002).

The validity of the sequence concept has been questioned on the basis
of deeper and larger blazar surveys (e.g. Giommi et al. 2005, Padovani
2007) which however lack until now the very important gamma-ray data.

Here we wish to address two points. The first concerns the validity of
the original claim {\it within the presently known bright blazar
SEDs}, the second concerns an anticipation of the types of blazars
that may be detected by GLAST.

\section{New data / new sources}
Given the limited space we will illustrate our points
schematically, commenting few
representative figures. All the figures will have in the background
the double humped lines interpolating the blazar spectral
sequence. The latter are just polinomial expressions connecting the
average monochromatic luminosities obtained as described above.

The SED of a new high redshift FSRQ serendipitously discovered by SWIFT 
(BAT) J0746+2548 (z=2.979) (Sambruna et al. 2006) is shown in Fig. 1a. 
Clearly J0746 is extremely luminous and conforms well to the sequence,
possibly suggesting a gamma-ray peak at Mev energies. 
The spectral shape in the gamma-ray band that will be measured by GLAST
for a large number of blazars will provide an essential information
to constrain the position of the high energy peak of blazar SEDs thus
probing the sequence concept.

\begin{figure}
  \includegraphics[height=.33\textheight]{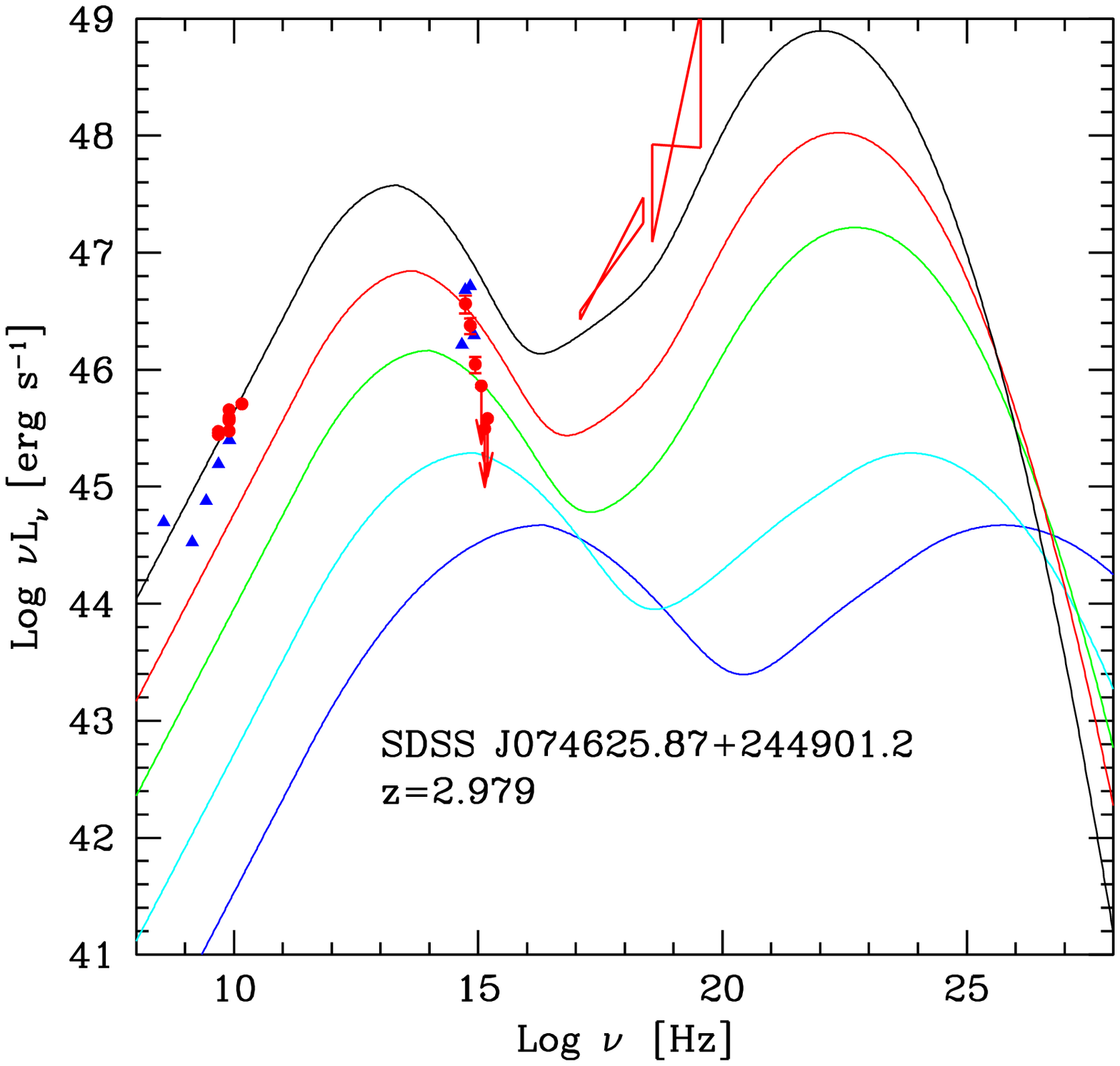}
  \includegraphics[height=.33\textheight]{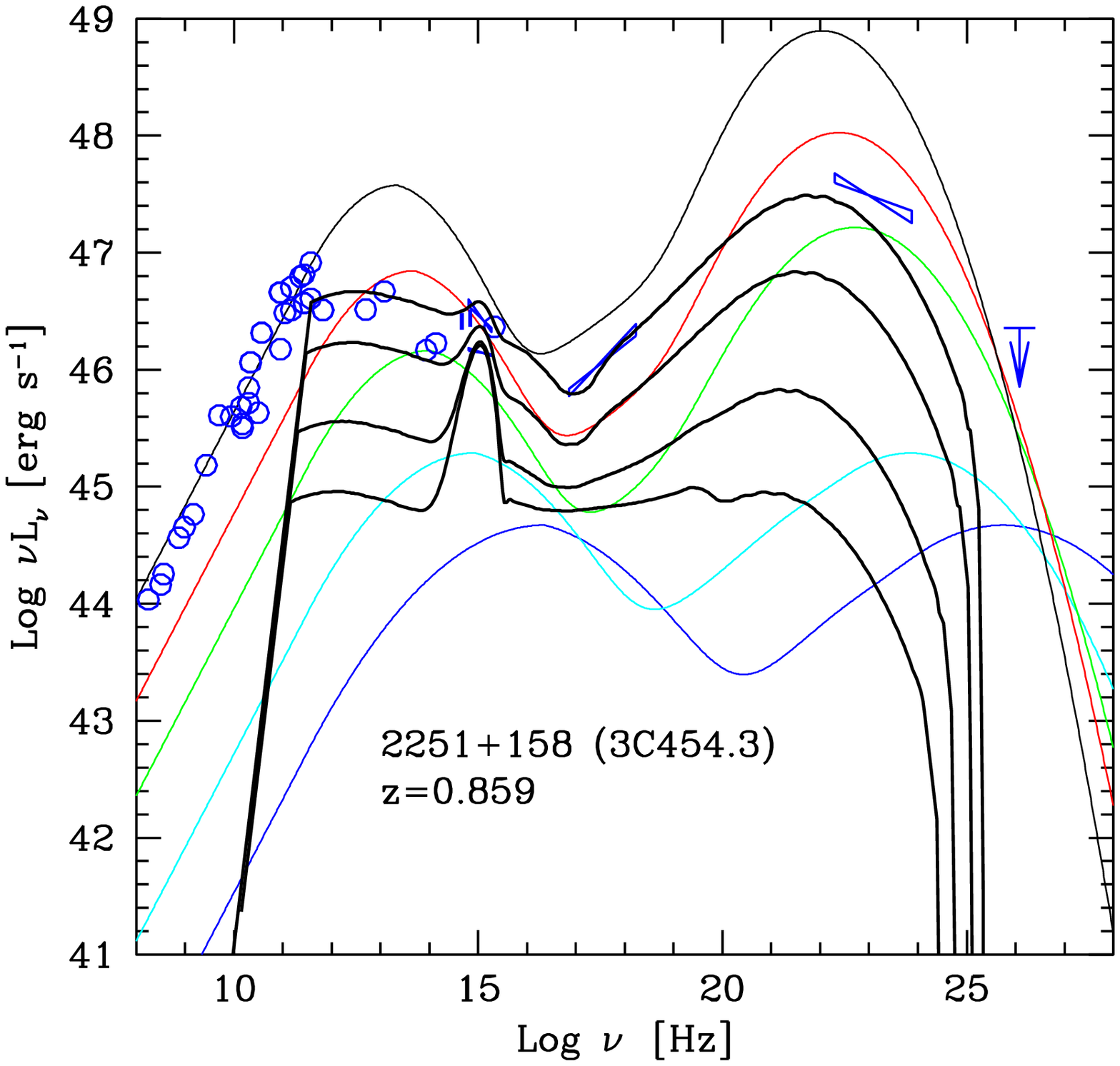}
  \caption{Spectral Energy Distribution of the blazars J0746+2548
  (left, from Sambruna et al. 2006) and 2251+158 (right, from
  Tavecchio et al. 2007) overimposed on the curves interpolating the blazar
  sequence. For 2251+158 we also report the model used to reproduce
  the data (upper black curve) and the emission expected for a
  misaligned jet with angles of respectively 6, 8 and 10 degrees
  (from top to bottom).}
\end{figure}

3C 454.3 is a highly variable FSRQ (z=0.859) already detected in
gamma-rays by EGRET.  The data for a "normal" state (Tavecchio et
al. 2007) are shown in Fig. 1b. This source could be detected with
GLAST at 1\% the intensity level shown in the figure which is the
average of EGRET measurements. The source underwent a strong outburst
recently and was observed by SWIFT (BAT) and INTEGRAL up to more than
100 keV (Pian et al. 2006, Giommi et al. 2006). In the latter state
the expected gamma-ray flux could have been an order of magnitude
brighter than detected by EGRET.  A source with an intrinsically
similar jet could then be detected in gamma-rays even if the jet was
at a larger angle to the line of sight.  The thick lines in Fig. 2
represent the model used to describe the "normal" state of 3C 454.3,
computed for different viewing angles. The gamma-ray emission could be
detected by GLAST up to an angle of 10 degrees to the jet axis. In
this case the SED would be significantly different than expected from
the sequence, simply because the jet emission is less beamed and less
prominent with respect the SED of the accretion disk, included here as
a blackbody component plus a Seyfert like X-ray component.
The sequence is not expected to extend to objects with jets seen at
intermediate angles. The different Doppler factor causes only a linear 
shift of the peak position but a dramatic change in luminosiy.  

Fig. 2 is devoted to blazars with lower luminosities. This part of the
sequence is populated exclusively by BL Lac objects defined as HBLs
due to their SEDs peaking at high energies, in the X-ray and TeV
bands.  In Fig. 2a the data for the "normal" state of PKS 2155-304 are
plotted in green.  They are well consistent with the sequence.  The
multifrequency data obtained during the exceptional TeV flare observed
from this source in July August 2006 are also shown (see Foschini et
al. 2007). During the outburst the two emission peaks do not appear to
shift much in frequency but the luminosities increase by a large factor
(for a short time) especially in the TeV band. Thus the high state SED
deviates remarkably from the sequence expectations.
For these objects, though relatively weak at GeV energies, GLAST
observations will be important to define the shape of the high energy
peak and  its possible evolution during outbursts. 

Finally, in Fig. 2b we show the data for 1629+4008 (z=0.272), a blazar
with an emission peak between the UV and the X-ray band discovered
within a survey aimed at finding objects with anomalous properties
(Padovani et al. 2002). The SED of this source complies reasonably
well with the sequence expectation for an HBL, however this object
shows emission lines which is not the case for HBLS. In fact the
sequence included only X-ray selected BL Lacs, but no X-ray selected
radio-loud objects with emission lines, as no such complete sample was
available at the time (see Wolter \& Celotti 2001).

This source indicates that jets with SEDs peaking at high energies can
occur in emission line AGNs. This is a new result, which however does
not break the correlations inferred from the sequence, as it occurs in
the low luminosity range. The question then is: what distinguishes
HBLs from objects like 1629? Why emission lines are completely absent
in HBLs but present in 1629 whose jet is of comparable luminosity?
According to our ideas (Maraschi 2001, Maraschi \& Tavecchio 2003) HBL
should accrete at highly subEddington rates, therefore in the
radiatively inefficient accretion (RIAF) regime, while 1629, which
shows emission lines, should be in the ``standard'' accretion disk
regime, therefore near to its Eddington limit. This in turn implies
that this source contains a central black hole of relatively modest
mass.  From the accretion luminosity, assuming that it corresponds to
0.1 the Eddington luminosity we can infer a mass of $6\times 10^7$
solar masses.  More direct estimates of the black hole mass are needed
to confirm this prediction.

\begin{figure}
  \includegraphics[height=.33\textheight]{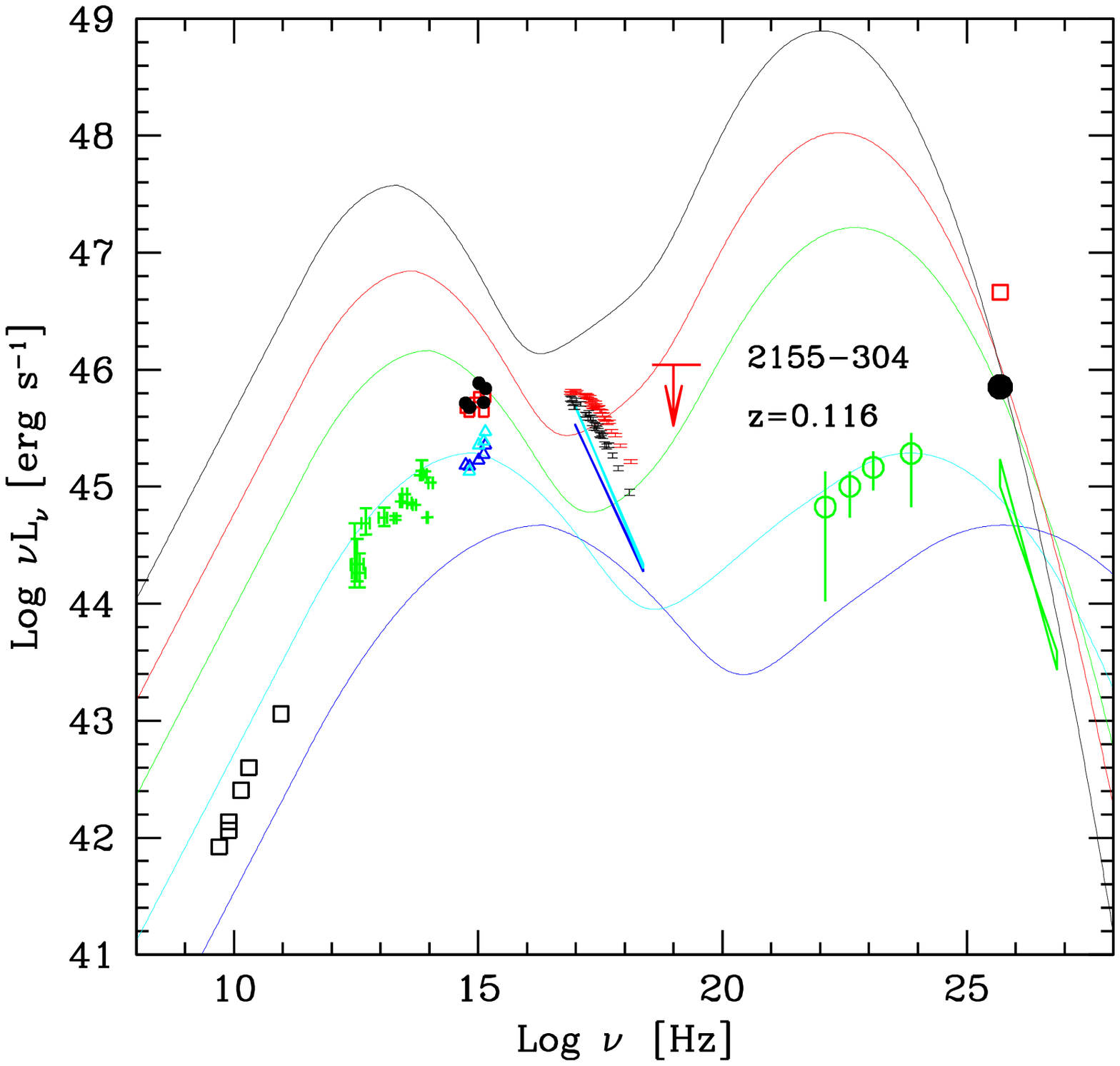}
  \includegraphics[height=.33\textheight]{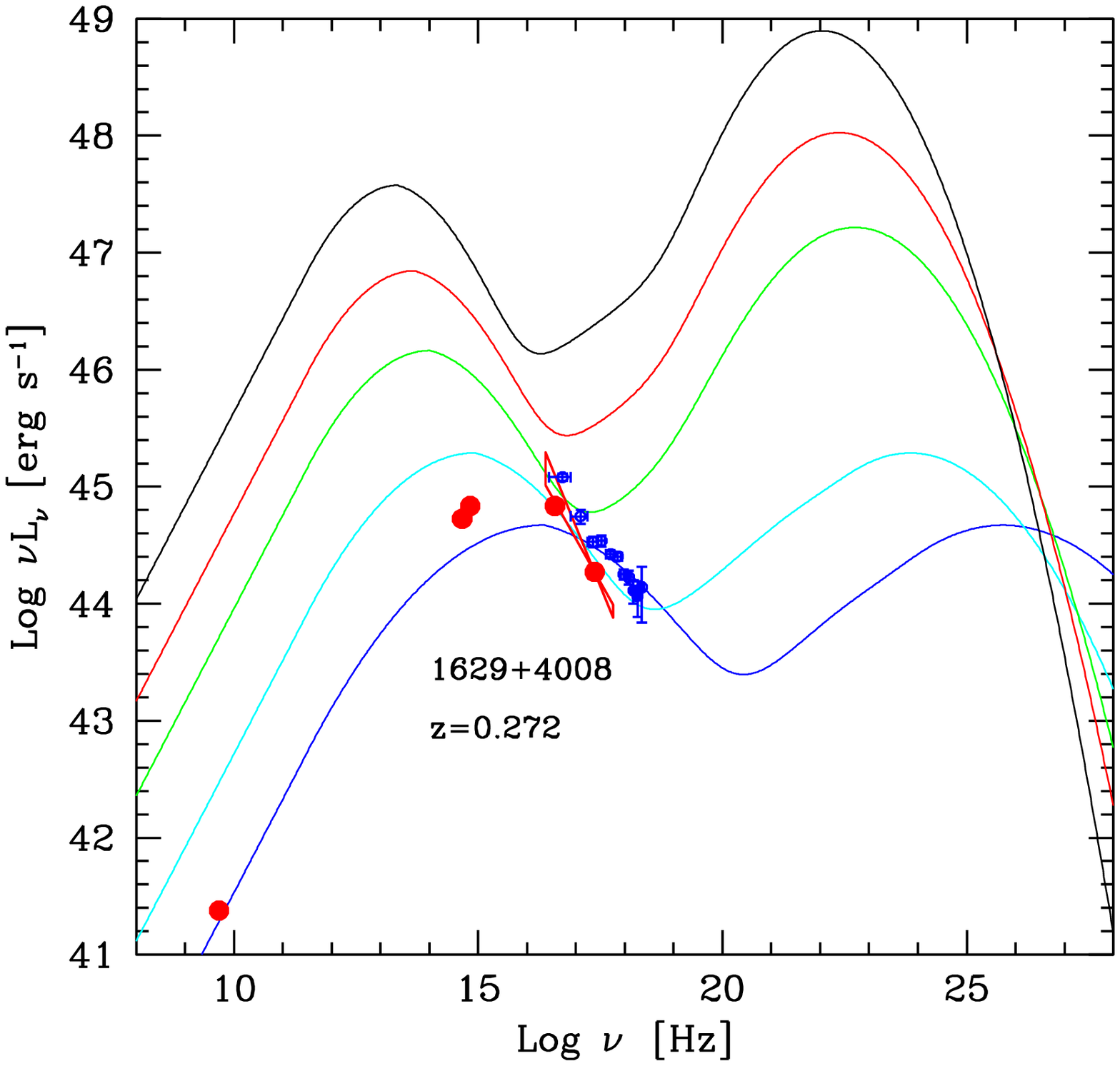}
  \caption{SEDs of the blazars 2251-304 (left, Foschini et al. 2007) 
  and 1629+4008 (Padovani et al. 2002) overimposed on the blazar sequence
interpolations. For PKS 2155-304 a normal state is shown together with 
optical/X-ray and TeV data during the exceptional outburst of July-August
2006}
\end{figure}

\section{Conclusions}

The few examples discussed above are meant to indicate how the
concept of a spectral sequence for blazars, based on averages over 
limited samples involving only the brightest objects of each class,
may be probed by GLAST. In particular, strong emphasis has
been put in the past on BL Lac objects, neglecting the X-ray selected
counterparts of FSRQ which may also be gamma-ray emitters. GLAST is
expected to produce extraordinary advances in this field.  It will
increase by orders of magnitude the number of objects with measured 
gamma-ray flux (see Dermer these proceedings) thus allowing to study 
deeper and differently selected samples. These will certainly 
contain "mixed"
objects in which the jet emission is less prominent in comparison to
other AGN properties. The new gamma-ray populations should carry great
potential for understanding the link between accretion power and the
production of jets in extragalactic objects.







\bibliographystyle{aipproc}   

\bibliography{sample}

\IfFileExists{\jobname.bbl}{}
 {\typeout{}
  \typeout{******************************************}
  \typeout{** Please run "bibtex \jobname" to optain}
  \typeout{** the bibliography and then re-run LaTeX}
  \typeout{** twice to fix the references!}
  \typeout{******************************************}
  \typeout{}
 }



\end{document}